\documentclass[twocolumn,showpacs,amsfonts,aps,prc,floatfix,%
superscriptaddress,nofootinbib]{revtex4-1} 

\usepackage{amsmath}
\usepackage{bm}
\usepackage{graphicx}

\newcommand{\be}[1]{\begin{equation}\label{#1}}
\newcommand{\ee}{\end{equation}}

\newcommand{\gsim}{\mbox{\raisebox{-0.6ex}{$\stackrel{>}{\sim}$}}\:}

\usepackage{color}

\begin{document}


\title{Longitudinal Viscous Hydrodynamic Evolution 
for the Shattered Colour Glass Condensate
}
\date{\today}

\author{Akihiko Monnai}
\email{monnai@nt.phys.s.u-tokyo.ac.jp}
\affiliation{Department of Physics, The University of Tokyo,
Tokyo 113-0033, Japan}

\author{Tetsufumi Hirano}
\email{hirano@phys.s.u-tokyo.ac.jp}
\affiliation{Department of Physics, The University of Tokyo,
Tokyo 113-0033, Japan}

\begin{abstract}
We investigate hydrodynamic evolution 
of the quark gluon plasma
for the colour glass condensate
type initial conditions.
We solve full second-order viscous hydrodynamic equations
in the longitudinal direction
to find that non-boost invariant expansion
leads to visible deformation on the initial
rapidity distribution. 
The results indicate that hydrodynamic evolution 
with viscosity plays an important role 
in determining parameters for the initial distributions.
\end{abstract}

\maketitle

The heavy ion program at Large Hadron Collider 
(LHC) 
in European Organization for Nuclear Research (CERN)
opens up new opportunities
to explore 
the deconfined matter, the quark gluon plasma (QGP) \cite{Yagi:2005yb},
in a wider temperature region.
There would be also a good opportunity
to investigate the colour glass condensate (CGC) \cite{Gelis:2010nm},
\textit{i.e.},
a universal form of colliding hadrons/nuclei
at very high energies,
of which we had a glimpse at Relativistic Heavy Ion Collider (RHIC)
in Brookhaven National Laboratory (BNL) \cite{Blaizot:2004px}.

Heavy ion reactions at high energies
undergo several stages such as collisions of two nuclei,
large entropy production just after the first contact,
local thermalisation,
hydrodynamic evolution and
chemical/thermal freezeouts of the system.
Since a framework to describe the CGC 
is regarded as an effective theory of high energy
hadrons/nuclei,
it is often employed to describe
the very first stage
in high energy heavy ion collisions
and to calculate
multiplicity and/or rapidity distributions
without assuming secondary interactions
\cite{
Kharzeev:2000ph,Kharzeev:2001gp,Kharzeev:2001yq,
Kharzeev:2004if,Albacete:2007sm,
Levin:2010zy,McLerran:2010ex}.
Nevertheless, the data are remarkably described
in this approach.
On the other hand, relativistic hydrodynamic models 
are quite successful to describe space-time evolution
of the QGP created in high energy heavy ion collisions,
in particular, anisotropy of transverse
collective flow \cite{HKH,Teaney:2000cw,Hiranov2eta,HT02}.
Initial states of the hydrodynamic
evolution, however, are still uncertain
so that quantitative conclusion about transport coefficients
depends on initial modeling \cite{Song:2010mg}.
This has been a longstanding issue
which should be by all means resolved
towards full understanding
of the QGP.
Both the initial gluon production
from the CGC and 
the hydrodynamic evolution of the QGP are
two distinct features of the whole reaction
so that it is indispensable to
unify these features  \cite{Hirano:2004en} and to
dynamically model the reactions as a whole from the
colliding two nuclei to final observables 
in a comprehensive fashion.

A vast body of relativistic ideal and viscous hydrodynamic 
simulations have been performed so far
to explore the bulk and transport properties of the QGP \cite{hydroreviews}.
Boost invariant expansion in the longitudinal direction \cite{Bjorken:1982qr}
is, however, often assumed in most of these simulations
to reduce the numerical efforts
even though boost invariant rapidity distributions
have never been observed.
Since particle production in
low transverse momentum region
is dominated by the small $x$
modes in the nuclear wave function,
where $x$ is a momentum fraction of
incident partons,
the QGP production could be traced back to
the initial parton density at small $x$ inside the colliding nuclei
before collisions.
The gluon density at small $x$ rapidly increases with decreasing $x$
and is eventually saturated
due to non-linear interactions among gluons.
The non-boost invariant
gluon production is predicted within the $k_{T}$-factorisation
formula
as a consequence of $x$ dependence of saturation scale, 
$Q_{s}^{2}(x) \propto x^{-\lambda}$, where 
$\lambda \sim$ 0.3 \cite{Gelis:2010nm,GBW}
controls rapidity and collision energy dependences of the saturation scale.

In this Letter we describe hydrodynamic evolution of the hot matter
in the beam direction
with initial conditions from the shattered CGC.
We focus especially on how hydrodynamic expansion
with or without viscosity
changes the flow-rapidity dependence of the entropy which could
be identified with final rapidity distribution of
hadrons.
We neglect transverse hydrodynamic expansion in this study.
Instead, we do not impose boost-invariance 
in the longitudinal expansion \cite{Chu:1986fu, Akase1989, Bozek:2007qt}.

We solve 
the full second order constitutive equations \cite{Monnai:2010qp}
generalised from the ones in the Israel-Stewart theory \cite{Israel:1979wp}
in the (1+1)-dimensional $\tau$-$\eta_{s}$ coordinates.
Here $\tau$ and $\eta_s$ are the proper time and the space-time rapidity,
respectively, defined as $t = \tau \cosh \eta_s$ and $z = \tau \sinh \eta_s$.
We neglect the baryon number current 
since we consider only gluon production in the $k_{T}$-factorisation formula
as a initial condition.
We choose the Landau frame 
where the energy dissipative current vanishes $W^\mu = 0$.
Then the constitutive equations are
\begin{eqnarray}
D \Pi &=& \frac{1}{\tau_{\Pi}} \bigg( -\Pi -\zeta_{\Pi \Pi} \frac{1}{T} \nabla Y_f - \zeta_{\Pi \delta e} D \frac{1}{T} \notag \\
&+& \chi_{\Pi \Pi}^b \Pi D \frac{1}{T} + \chi_{\Pi \Pi}^c \Pi \nabla Y_f + \chi_{\Pi \pi} \pi \nabla Y_f \bigg) ,
\label{eq:Pi} 
\end{eqnarray}
\begin{eqnarray}
D \pi &=& \frac{1}{\tau_{\pi}} \bigg( -\pi + \frac{4}{3} \eta \nabla Y_f 
+ \chi_{\pi \pi}^b \pi D \frac{1}{T} + \chi_{\pi \pi}^c \pi  \nabla Y_f  
\notag \\
&+& \frac{2}{3} \chi_{\pi \pi}^d \pi \nabla Y_f + \frac{2}{3} \chi_{\pi \Pi} \Pi \nabla Y_f \bigg) ,
\label{eq:pi}
\end{eqnarray}
where $\Pi$ is the bulk pressure and $\pi$ 
the shear pressure defined with the shear stress tensor $\pi^{\mu \nu}$
as $\pi = \pi^{00} -\pi^{33}$. 
Note that, in the (1+1)-dimensional geometry,
it is sufficient to treat 
$\pi$ as the only independent component
from the orthogonality and the traceless conditions.
$T$ is the temperature and $Y_f$ is the flow rapidity defined by
the four-fluid velocity $u^\mu = (\cosh{Y_f},0,0,\sinh{Y_f})$.
$\eta$ is the shear viscosity, $\zeta_{\Pi \Pi}$ and $\zeta_{\Pi \delta e}$ the bulk viscosities, $\tau_\Pi$ and $\tau_\pi$ the relaxation times and $\chi_{\Pi \Pi}^b$, $\chi_{\Pi \Pi}^c$, $\chi_{\Pi \pi}$, $\chi_{\pi \pi}^b$, $\chi_{\pi \pi}^c$, $\chi_{\pi \pi}^d$ and $\chi_{\pi \Pi}$ the second order transport coefficients.
Here we have two ``bulk viscosities",
$\zeta_{\Pi \Pi}$ and $\zeta_{\Pi \delta e}$,
as required in the linear response theory.
See also Ref.~\cite{Monnai:2010qp}.
The time- and the space-like derivatives in this geometry are
$
D = \cosh (Y_f -\eta _s) \partial _\tau + 
\frac{1}{\tau} \sinh (Y_f -\eta _s) \partial _{\eta _s} \label{D}$and 
$
\nabla = \sinh (Y_f -\eta _s) \partial _\tau + 
\frac{1}{\tau} \cosh (Y_f -\eta _s) \partial _{\eta _s} .\label{K} $
Hydrodynamic equations
become complicated
in non-boost invariant case $Y_{f} \neq \eta_{s}$
because derivatives with respect to the proper time and 
the space-time rapidity are mixed, which 
significantly increases the numerical difficulty
compared with the constitutive equations
in the transverse plane assuming the boost invariant flow.

We solve the constitutive equations (\ref{eq:Pi}) and (\ref{eq:pi}) 
together with the energy-momentum conservation equations
in the piecewise parabolic method \cite{PPM} 
which was employed in one of the most successful
(3+1)-dimensional ideal hydrodynamic calculations \cite{Hiranov2eta}.
The numerical difficulty raised 
by the mixing of the derivatives
with respect to $\tau$ and $\eta_{s}$
is dealt with by taking iteration on the expansion scalar
$\theta =  \nabla Y_f$.
We have checked the solutions converge typically in several steps.
Numerical details will be discussed elsewhere \cite{MH}.

One needs the equation of state $P_0=P_0(e_0)$ and 
the transport coefficients,
which contain microscopic information of the systems,
to perform hydrodynamic simulations.
We employ 
the latest (2+1)-flavor lattice QCD result \cite{Borsanyi:2010cj}
for the equation of state.
On the other hand, we introduce some models
for the transport coefficients
since, to our knowledge,
there is no single framework
which gives all the transport coefficients appeared in this study.
Here
we use the conjectured minimum bound for the ratio of shear viscosity 
to the entropy density
$\eta/s = 1/{4\pi}$ from Anti-de Sitter/conformal field theory
(AdS/CFT) correspondence \cite{Son_visc}
just for the purpose of demonstration
to see how visible the entropy distribution changes
during the longitudinal evolution.
On the other hand, there are two bulk viscous coefficients
$\zeta_{\Pi \Pi}$ and $\zeta_{\Pi \delta e}$
when the first-order cross terms are properly kept.
So far there have been few calculations on these coefficients.
Hence we try to get an insight for these two coefficients
from the $\phi^4$-theory
in the non-equilibrium statistical operator method \cite{Hosoya:1983id}
and calculate the ratios $\zeta_{\Pi \Pi}/\eta$
and $\zeta_{\Pi \delta e}/\eta$ as $5T/6$ and $-5T/2$, respectively. 
Using the energy-momentum conservation and the Gibbs-Duhem relation,
the two linear terms are merged into one term at the first order as
\begin{eqnarray}
- \zeta_{\Pi \Pi} \frac{1}{T} \nabla_\mu u^\mu 
- \zeta_{\Pi \delta e} D \frac{1}{T} 
= - \frac{\zeta_{\Pi \Pi}  
+ c_s^2 \zeta_{\Pi \delta e}}{T} \nabla_\mu u^\mu ,
\end{eqnarray}
where $\zeta = (\zeta_{\Pi \Pi}  + c_s^2 \zeta_{\Pi \delta e})/T
= \frac{5}{2} ( \frac{1}{3} - c_s^2 ) \eta$
corresponds to the conventional bulk viscous coefficient.
The relaxation times $\tau_\Pi$ and $\tau_\pi$ and 
the other second order transport coefficients $\chi_{\Pi \Pi}^b$, 
$\chi_{\Pi \Pi}^c$, $\chi_{\Pi \pi}$, $\chi_{\pi \pi}^b$, 
$\chi_{\pi \pi}^c$, $\chi_{\pi \pi}^d$ and $\chi_{\pi \Pi}$ are 
estimated in kinetic theory as in Ref.~\cite{Monnai:2010qp}.
The coefficients estimated with hadronic components 
up to $\Delta (1232)$ are connected to those with $u, d, s$ quarks
 and gluons as its components by hyperbolic factors
 around the (pseudo-)critical temperature 
as $\chi = \frac{1}{2}(1- \tanh \frac{T-T_0}{\Delta T}) \chi_\mathrm{had}
 + \frac{1}{2}(1+ \tanh \frac{T-T_0}{\Delta T}) \chi_\mathrm{udsg}$
where $T_0=0.17$ GeV and $\Delta T = 0.02$ GeV.
The relaxation times are calculated likewise.
We emphasize here that they are trial parameters to investigate qualitative response of the hot matter and that obtaining realistic transport coefficients is not the aim of this Letter.

The initial conditions for hydrodynamic simulations are
obtained from Monte Carlo
version \cite{Drescher:2006ca}
of $k_T$-factorisation formula 
with unintegrated gluon distributions parametrised
by Kharzeev, Levin and Nardi (MC-KLN) \cite{Kharzeev:2001gp,Kharzeev:2002ei}. 
In this model the saturation scale $Q_s$
for a nucleus $A$ at a transverse coordinate $\bm{x}_\perp$ is given as 
\begin{eqnarray}
Q_{s,A}^2(x;\bm{x}_\perp) = Q_{s,0}^2 \frac{T_A (\bm{x}_\perp)}{T_{A,0}}
\bigg( \frac{x_0}{x} \bigg)^\lambda ,
\end{eqnarray}
where $T_A (\bm{x}_\perp)$ is a thickness function
which is obtained from randomly distributed hard source (nucleons)
according to the Woods-Saxon nuclear profile.
We use the same parameter set, 
$Q^2_{s,0} = 2$ GeV$^2$, $T_{A,0} = 1.53$ fm$^{-2}$, $\lambda = 0.28$
and $x_{0} = 0.01$
as in Ref.~\cite{Hirano:2009ah}.
The 5\% most central events in the Monte-Carlo
calculations are employed
for constructing the smooth initial conditions.
The initial energy density distribution 
as a function of space-time rapidity, $e_{0}(\tau_{0},\eta_{s})$ 
is obtained
from the transverse energy distribution
$dE_T/dy$ over the overlapping area of
the nuclei $S_\mathrm{area}$
by identifying momentum rapidity $y$ with space-time rapidity $\eta_s$.
Although the pre-thermalisation stage
would be very complicated, we simply assume,
during a very short period until the thermalisation time
$\tau_0 = 1$ fm$/c$,
the energy density in a fluid element at $\tau_0$
is the same as
locally deposited energy $dE/d^2x_\perp \tau_{0} d\eta_s$ 
calculated in the CGC.
Maximum (minimum) $p_T$ cut is set to 3 (0.1) GeV/$c$
to include the contribution 
from low $p_T$ partons which is assumed 
to form thermally-equilibrated media.
When the saturation scale is smaller than $\Lambda_{\mathrm{QCD}}$,
we simply assume gluons are not produced in $k_T$-factorisation
formula.
This prescription leads to reduction of the gluon
multiplicity at mid-rapidity
by 21.5\% at RHIC and 39.9\% at LHC.
The temperatures at mid-rapidity at the initial time are 419 MeV for RHIC
and 490 MeV for LHC in these settings.
We assume the boost-invariance 
only for the initial flow, \textit{i.e.}, $Y_f (\tau_0) =\eta_s$,
not for any other hydrodynamic initial variables.
It is more difficult to determine 
the initial conditions for 
the shear pressure $\pi$ and the bulk pressure $\Pi$
because the precise dynamics 
before the local thermalisation is not fully known.
Here we set them to be vanishing since assuming finite
dissipative currents at $\tau_{0}$
corresponds to employing different initial
energy-momentum tensors between ideal and viscous cases,
which would make a comparison of the results and
estimation of the viscous effects difficult.

We investigate the entropy distribution 
as a function of the flow rapidity $dS/dY_f$
which roughly corresponds to the rapidity distribution
of hadrons \cite{Morita:2002av}. 
This can be interpreted as follows.
The entropy density $s$ carries the information 
of the number density in relativistic massless ideal gas limit
since the ratio between the two quantities should be temperature
independent as indicated by a dimensional analysis.
Momentum rapidity $y$, on the other hand,
can be identified with the flow rapidity $Y_{f}$
in a fluid element on average.
Thus, this is the closest quantity to the rapidity distribution
we have from a pure hydrodynamic point of view,
which do not include complicated freezeout processes nor
any other model assumptions.
Since the second order viscous correction
to $s^\mu$ should not be large,
we discuss here the lowest order modification.
One can make a rough estimation of charged particle
rapidity density by $dN_{\mathrm{ch}}^{\mathrm{hydro}}/dy \approx (2/3)\times (1/3.6) \times dS/dY_{f}$.

\begin{figure*}[htb]
\includegraphics[width=3.4in]{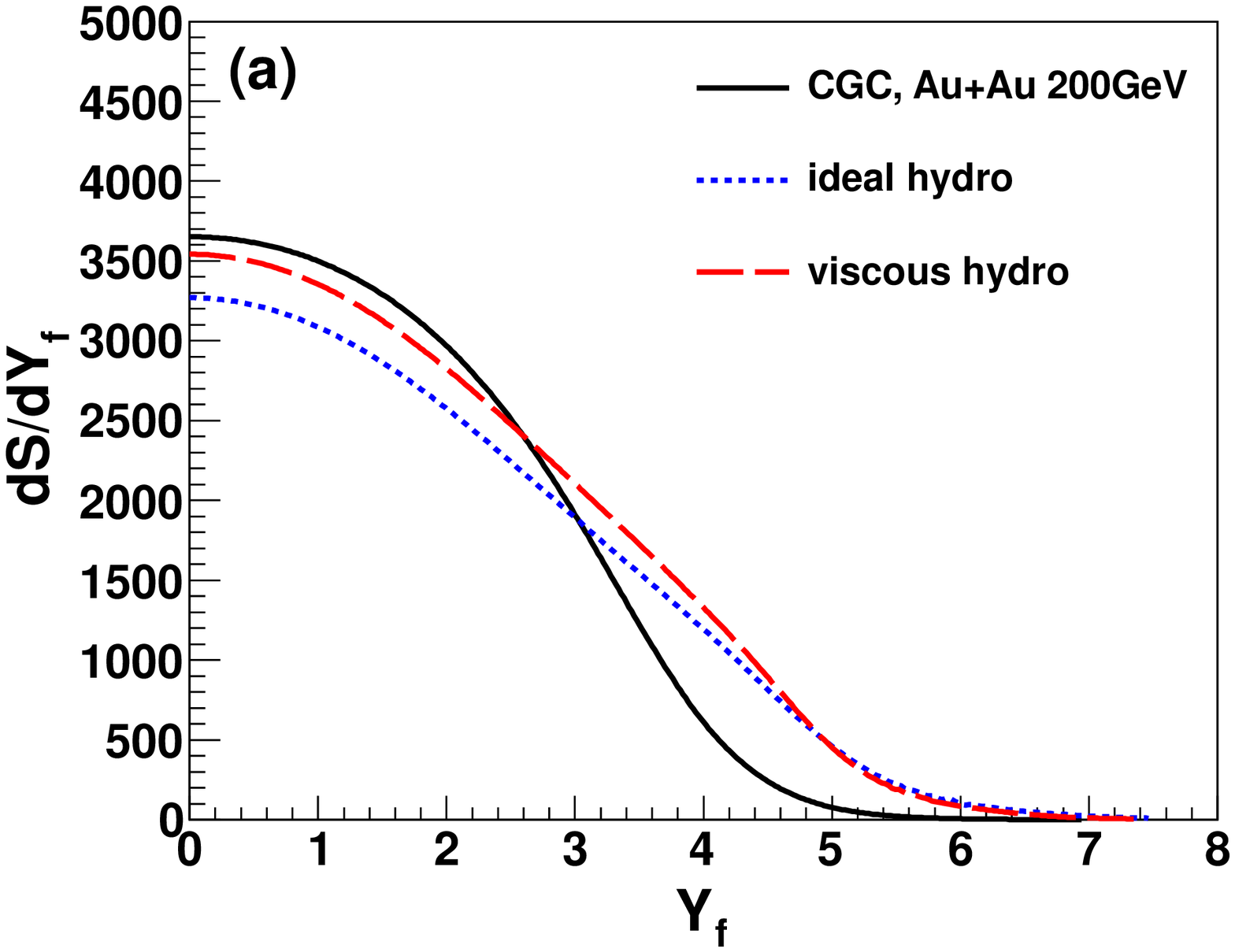}
\includegraphics[width=3.4in]{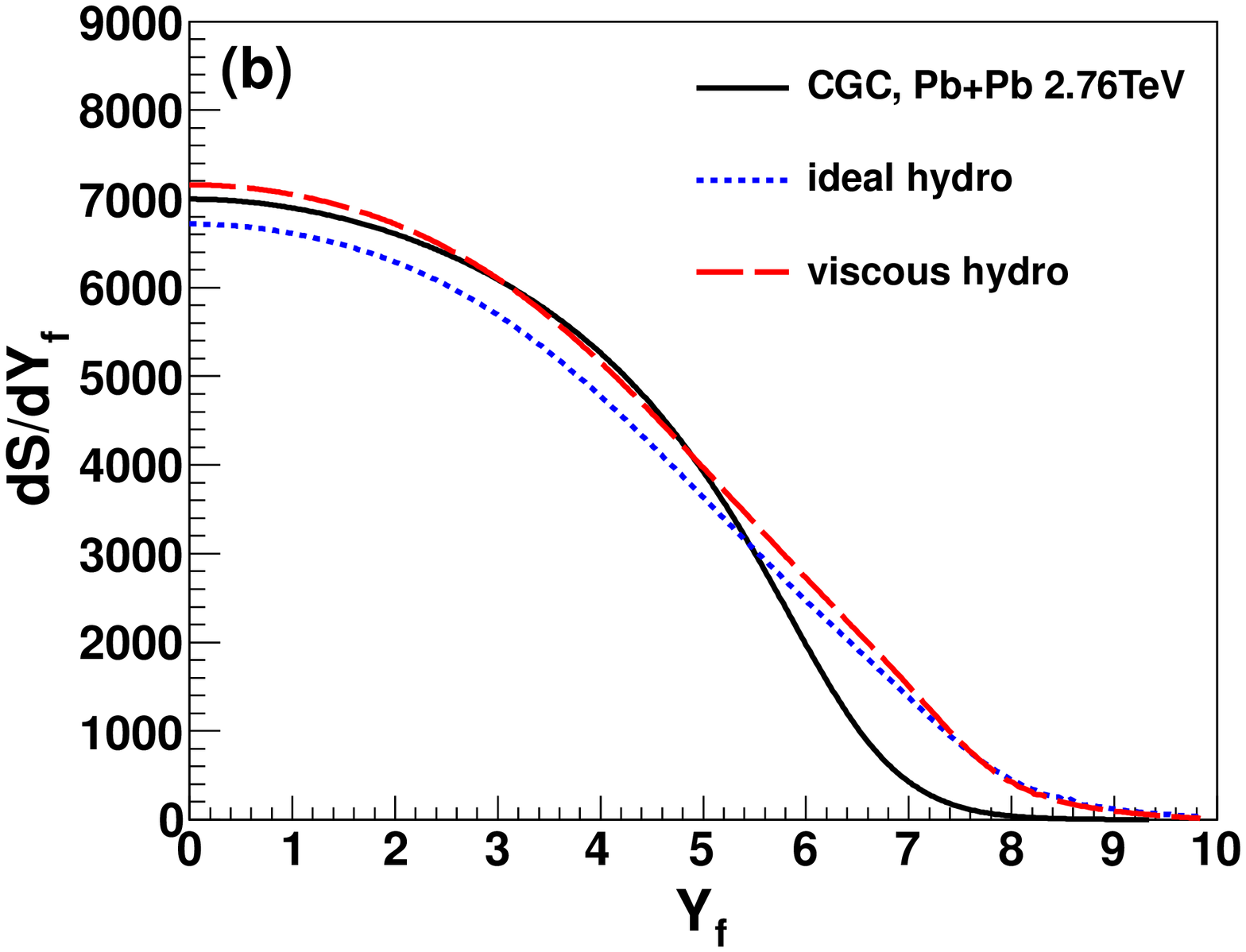}
\caption{(Colour online) 
The initial $dS/dY_f$ distributions at $\tau = 1$ fm/$c$
from the colour glass condensate (solid line)
and the final distributions after the ideal hydrodynamic (dotted line)
and the shear and bulk viscous hydrodynamic (dashed line) evolution for
(a) Au+Au collisions at RHIC ($\tau = 30$ fm/$c$)
and (b) Pb+Pb collisions at LHC ($\tau = 50$ fm/$c$).
}
\label{fig:dsdyf}
\end{figure*}

The flow rapidity dependences of entropy 
are shown in Fig.~\ref{fig:dsdyf} for
Au+Au and Pb+Pb collisions at $\sqrt{s_{NN}} =$ 
200 GeV and 2.76 TeV, respectively.
The final times are $\tau = 30$ fm/$c$ and
$\tau = 50$ fm/$c$ for the Au+Au and the Pb+Pb cases, respectively.
These are the typical times at which 
the temperature at mid-rapidity is sufficiently close
to the (pseudo-)critical temperature $T_c \sim 0.17$ GeV. 
Note that they are much longer than 
a conventional lifetime of the QGP in the heavy ion collisions
because the transverse expansion neglected in the present study
would accelerate the cooling process.
%

One can see that the hydrodynamic evolution
visibly modifies the initial distributions
from the CGC model to more flattened ones at both RHIC and LHC energies. 
This behaviour can be seen only
if boost-invariance is not assumed, \textit{i.e.},
a pressure gradient with respect to $\eta_{s}$
exists.
Since entropy is produced 
in non-equilibrium hydrodynamic evolution,
the entropy distributions are
larger than those of the ideal hydrodynamic systems
in most of the flow rapidity region.
Final entropy at mid-flow rapidity
results from the interplay between the entropy production and outward flow
which carries entropy to the forward rapidity region.
The ideal hydrodynamic process always lowers the yield of 
the initial distribution at mid-flow rapidity due to the outward flow.
The viscous hydrodynamic one gives rather non-trivial results
since the distribution is still lowered for 
the 200 GeV Au+Au collisions,
but is enhanced for the 2.76 TeV Pb+Pb collisions.
The difference is not due to the different choices
of the final times, because, as we will see later,
the shapes of the distributions do not change
so much after $\tau \sim 20$~fm/$c$ for both cases.
The modification of the initial distribution
is rather sensitive to the shape of the initial distribution.
It should be noted that the corrections in the larger 
$Y_f$ region ($Y_f \gsim 3$ at RHIC and $Y_f \gsim 6$ at LHC)
might be slightly overestimated
because even though the distribution shapes
steady at a relatively early stage,
the temperature in the forward region still
hits $T_c$ earlier than the mid-rapidity regions do
and there would be additional hydrodynamic correction
at the final times.
This, of course, does not change the conclusion that the yield
at the smaller rapidity regions are affected by hydrodynamic evolution.
Considering that we use small shear and bulk viscous coefficients
close to the conjectured minimum boundaries \cite{Son_visc, Buchel:2007mf} 
in the calculation,
the results also suggest that effects of viscosity would be important.
It should be noted that the numerical results here include
only the longitudinal expansion
and that the transverse expansion, which is missing in the present study,
would be important in more quantitative discussion.

The fact that the initial entropy distribution $dS/dY_{f}
\approx dS/dy \propto dN/dy$
could be deformed during the hydrodynamic stage
is of particular importance
since 
initial gluon rapidity distribution 
is often directly compared with the observed
charged hadron data assuming the parton-hadron duality
\cite{
Kharzeev:2000ph,Kharzeev:2001gp,Kharzeev:2001yq,
Kharzeev:2004if,Albacete:2007sm,
Levin:2010zy,McLerran:2010ex}.
As we saw above, the hydrodynamic expansion beyond
the boost invariant flow could make a change of initial rapidity
distribution.
Thus, the $\lambda$ parameter in some CGC-inspired models
is subject to correction due to hydrodynamic
effects if one wants to constrain it through
the rapidity distribution.
Even if one satisfactorily reproduced
the rapidity distribution at RHIC
with a set of CGC parameters assuming parton-hadron duality,
one could fail to reproduce LHC data with the same parameter set
due to lack of hydrodynamic corrections. 
Suppose the result from the viscous hydrodynamic
model in Fig.~\ref{fig:dsdyf}~(a)
would reproduce hadron rapidity distribution at RHIC,
then
the actual $\lambda$ which includes the 
hydrodynamic expanding effects would be larger than
an apparent $\lambda$ which can be obtained by fitting 
this rapidity
distribution without considering any secondary scatterings, 
as the distribution tends to become
steeper as increasing $\lambda$.
If one assumes that the hydrodynamic effect 
is smaller at LHC as indicated in Fig.~\ref{fig:dsdyf}~(b),
this could be one of the possible interpretations
for the fact that most of CGC models
turned out to underpredict the multiplicity in the Pb+Pb collisions
at LHC \cite{Aamodt:2010pb}.
Since the entropy production due to viscosity
is non-trivial as increasing collision energy, 
the energy dependence of multiplicity even at mid-rapidity
from the CGC is also subject to hydrodynamic correction. 

\begin{figure*}[htb]
\includegraphics[width=3.4in]{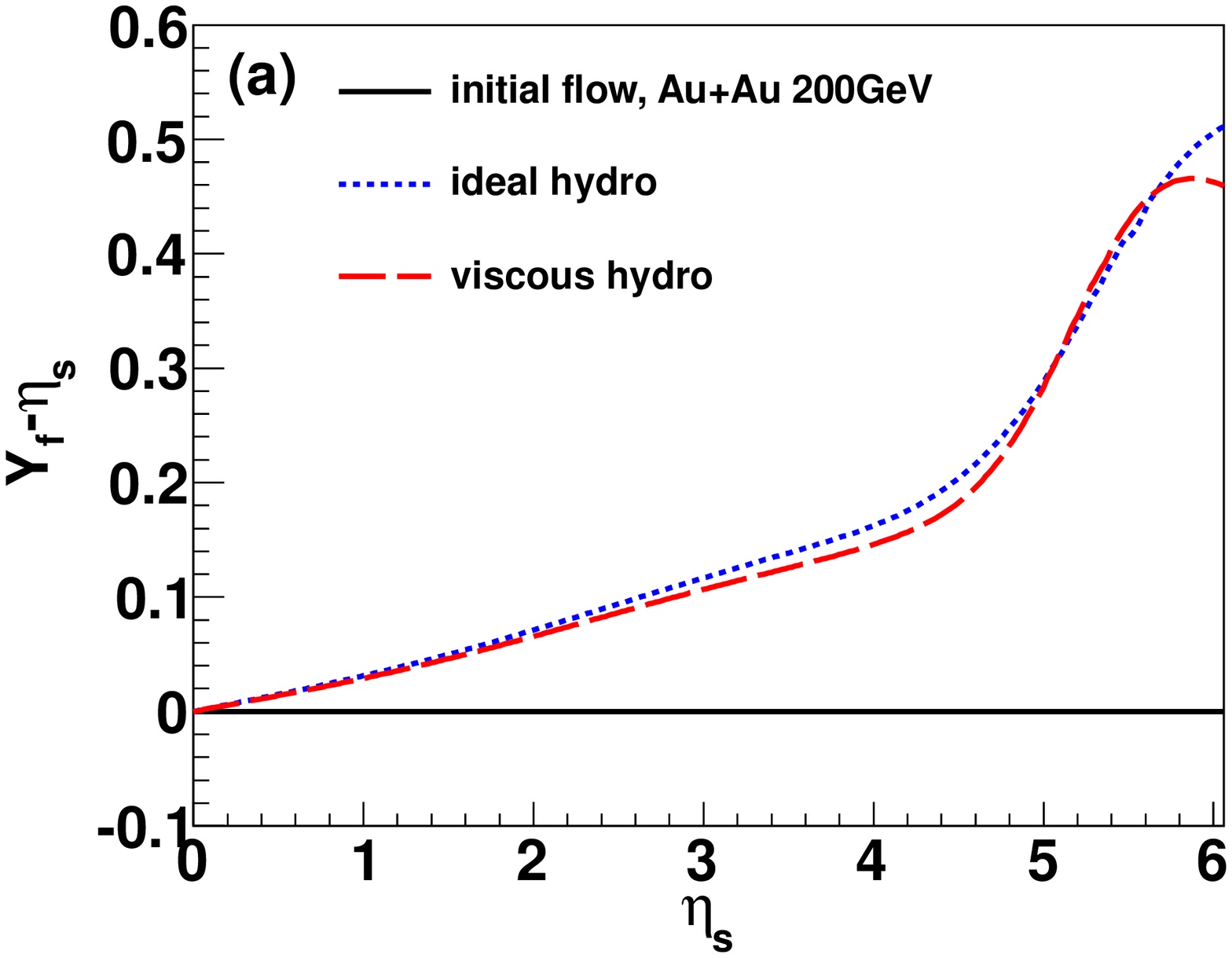}
\includegraphics[width=3.4in]{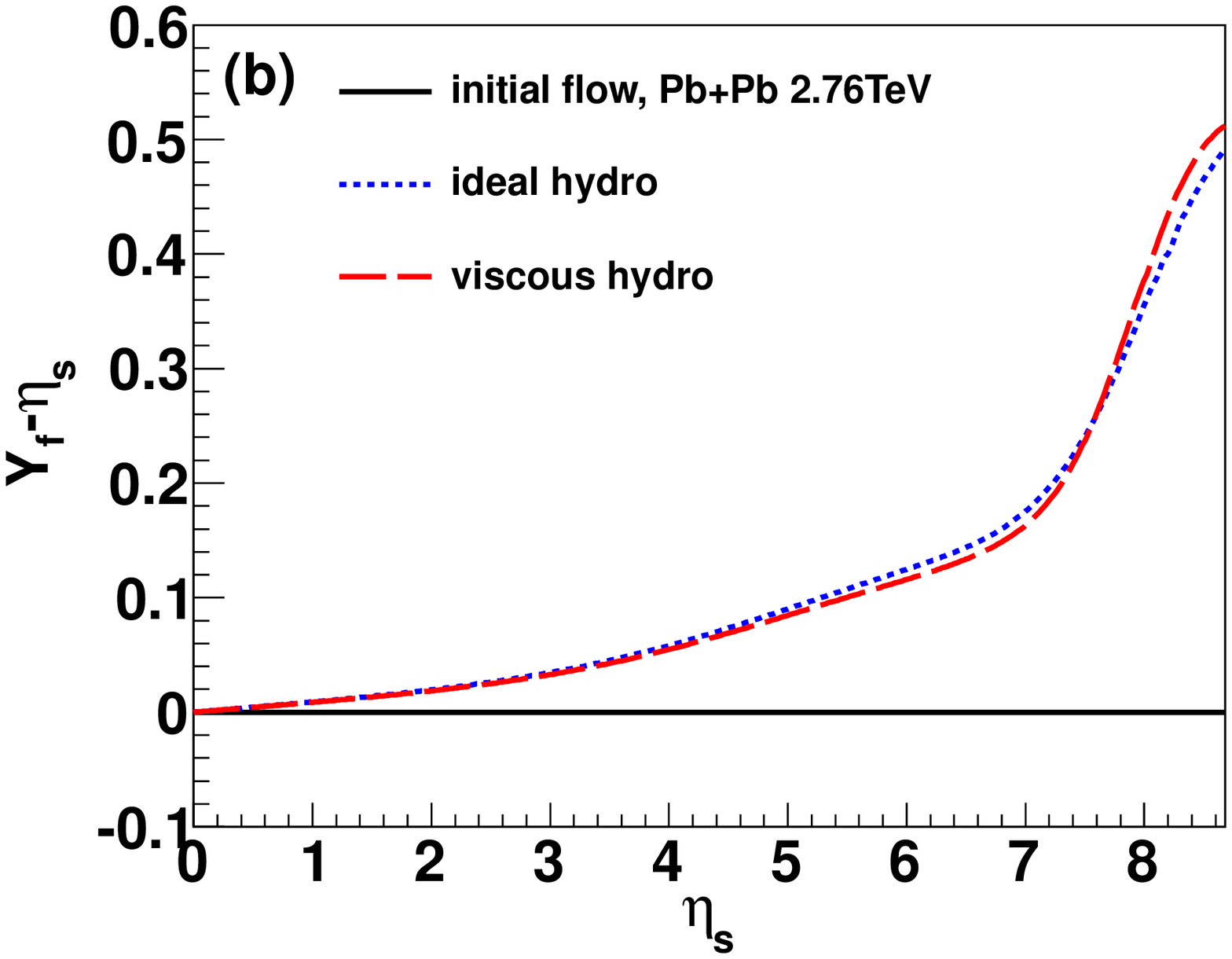}
\caption{(Colour online) The deviation of the flow rapidity
from the space-time rapidity
$Y_f - \eta_s$ at the initial time $\tau = 1$ fm/$c$ (solid line)
and after the ideal hydrodynamic (dotted line)
and the shear and bulk viscous hydrodynamic (dashed line)
evolution for (a) Au+Au collisions at RHIC
($\tau = 30$ fm/$c$) and (b) Pb+Pb collisions at LHC ($\tau = 50$ fm/$c$).
}
\label{fig:yfetas}
\end{figure*}

To further quantify the deviation from
the boost invariance ($Y_f=\eta _s$), 
we also estimate the difference
between the flow rapidity and the space-time rapidity,
$Y_f-\eta _s$.
Results for Au+Au collisions at $\sqrt{s_{NN}} = 200$
GeV and Pb+Pb collisions at $\sqrt{s_{NN}} = 2.76$ TeV
are shown up to the beam rapidities in Fig.~\ref{fig:yfetas}.
%
One sees in all cases that the deviations
from the boost-invariant flow are positive
and become large towards larger space-time rapidity
due to the acceleration by the pressure gradient in the beam direction.
Both Au+Au case at RHIC and Pb+Pb case at LHC 
exhibit the same trend, while the latter is slightly moderate
near mid-rapidity
possibly because initial energy density profile is less steep in the case. 
The deviations for the viscous cases are slightly smaller
than those for the ideal cases for the $Y_f$ regions
in which the fluids are relatively hot,
because the pressure gradients
are effectively reduced in the longitudinal direction
by the bulk and the shear pressures
at early times in the space-time evolution.
The situation, however, is different
for the later times
because the entropy generated in the viscous systems
enhances the pressure gradients, 
while the corrections from the shear and the bulk pressures 
themselves are already small.
Due to the counter contributions at the late stage,
the overall differences between the ideal and the viscous flow deviations
are small at those proper times.

Finally, we demonstrate the time evolution of
$dS/dY_f$ and $Y_f - \eta_s$
in Pb+Pb collisions 
at $\sqrt{s_{NN}}=$ 2.76 TeV. 
\begin{figure*}[htb]
\includegraphics[width=3.4in]{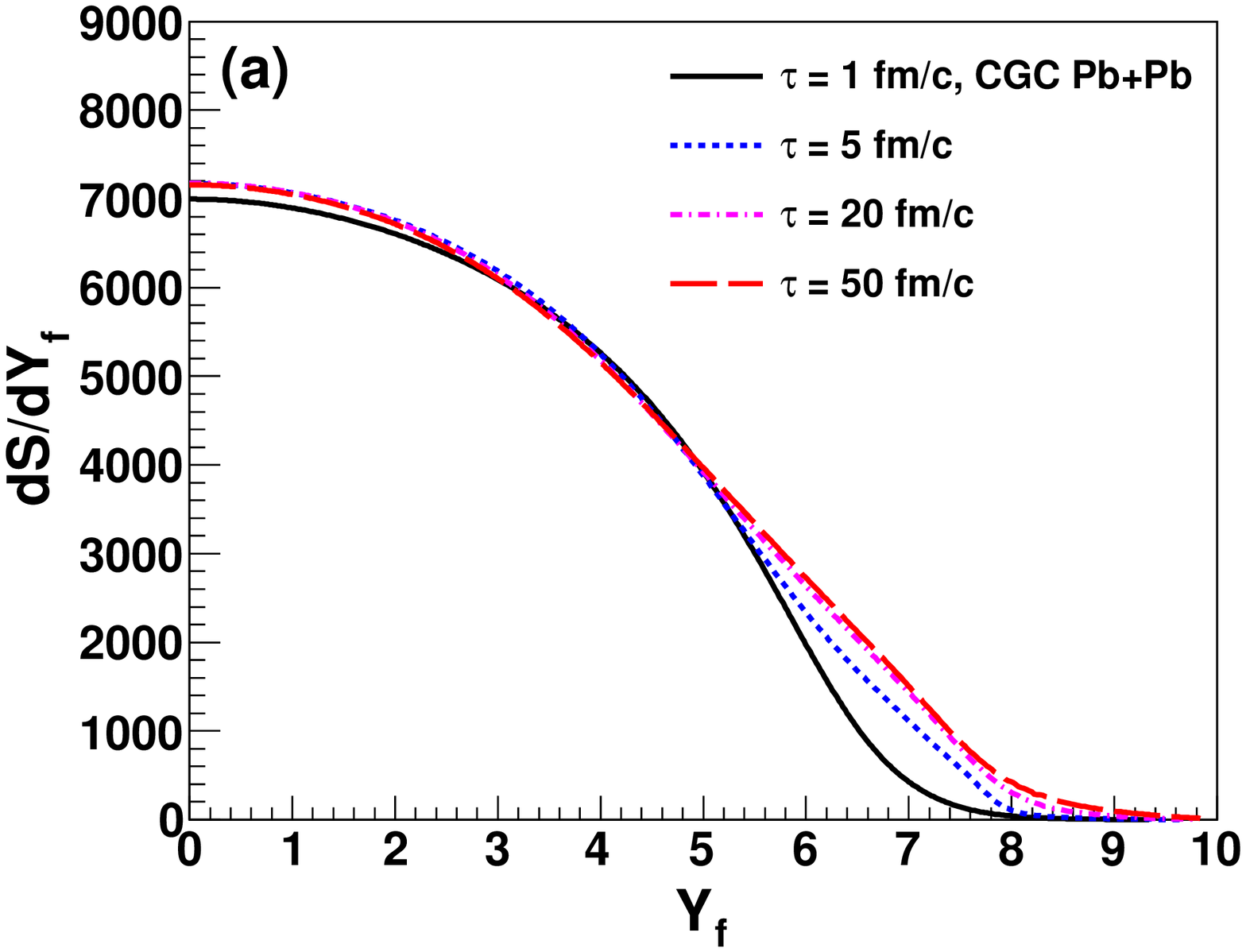}
\includegraphics[width=3.4in]{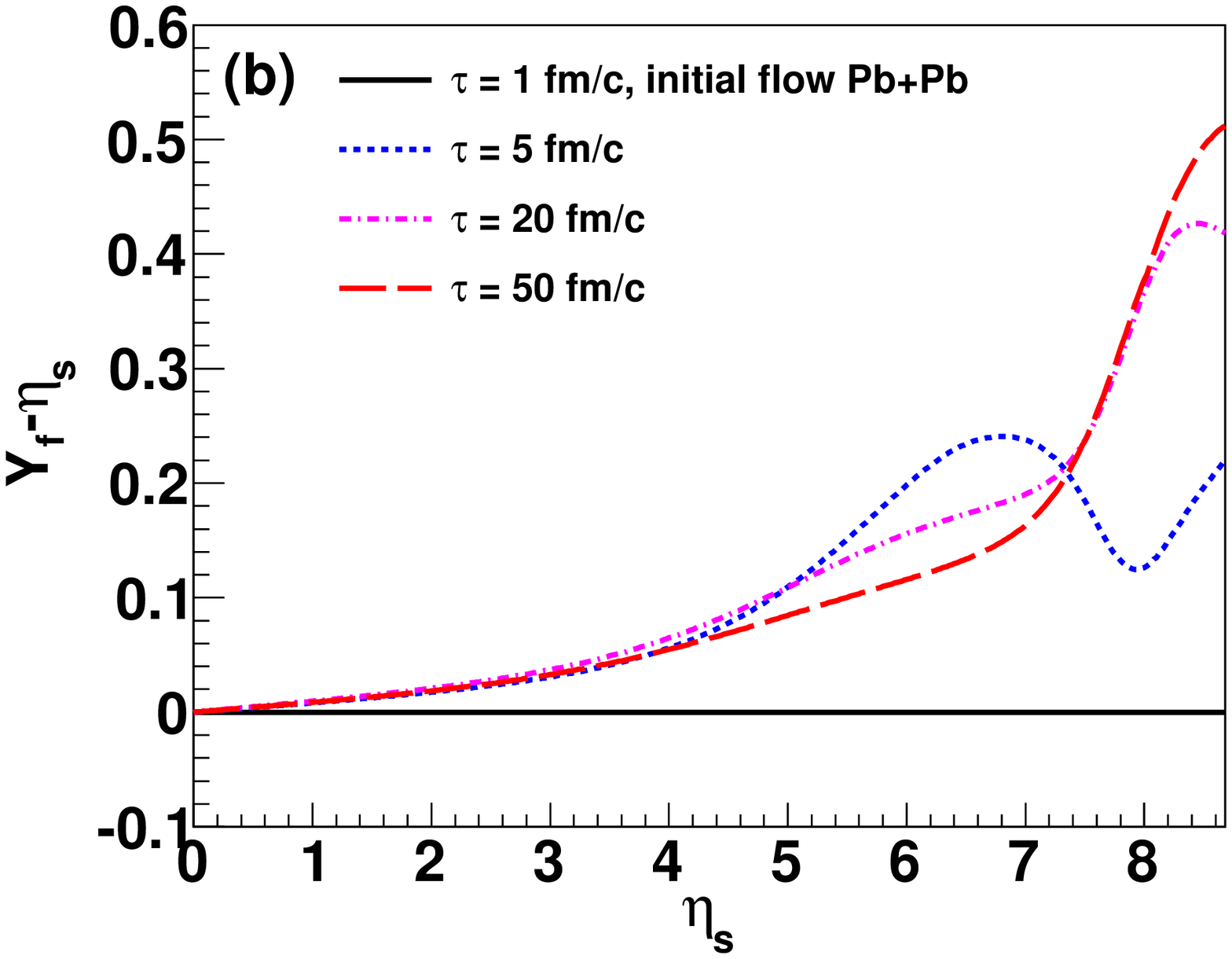}
\caption{(Colour online) (a) The deformation of the initial
entropy distribution per flow rapidity and
(b) the deviation of the flow rapidity from the space-time rapidity
$Y_f - \eta_s$ at initial time $\tau = 1$ fm/$c$ (solid line),
$\tau = 5$ fm/$c$ (dotted line),
$\tau = 20$ fm/$c$ (dash-dotted line)
and $\tau = 50$ fm/$c$ (dashed line)
in viscous hydrodynamic evolution in Pb+Pb collisions at the LHC energy.
}
\label{fig:t}
\end{figure*}
%
In Fig.~\ref{fig:t} (a), the proper time dependence of
the entropy distribution per flow rapidity is shown
at the time $\tau = 1$, 5, 20 and $50$ fm/$c$.
As mentioned earlier,
the entropy distribution in the hydrodynamic
evolution does not change much its shape after
$\tau \sim 20$ fm/$c$.
The yield around $Y_{f} = 0$ is  
almost constant throughout the time evolution.
It is due to a rather accidental cancellation between the entropy production
and the expansion by the outward entropy flux in the parameter settings.
This means we have a monotonous decrease at mid-flow rapidity
in the case of an ideal hydrodynamic calculation.
In Fig.~\ref{fig:t} (b),
one sees the dynamical evolution of the deviation
from the boost-invariant flow. At $\tau = 5$ fm/$c$,
there are an rise and a dip in the flow acceleration near
$Y_f \sim$ 6-7 and 7-8, respectively, 
because the effective pressure
$P_0 + \Pi - \pi$
can become very small 
in the large $Y_f$ region
when the absolute values of
$\Pi$ and $\pi$ are still large.
The sudden decrease in pressure leads to 
an enhancement in its gradient followed by a suppression.
Note that the effect quickly disappears as 
the dissipative currents rapidly approach vanishing
along with the time evolution.
Eventually the flow rapidity distribution evolves
into the one closer to the ideal hydrodynamic distribution
we have seen in Fig.~\ref{fig:yfetas}.
It is worth-mentioning that, unlike $dS/dY_f$,
the flow rapidity profile changes
in the time evolution after $\tau = 20$ fm/$c$ for the current parameter sets. 

To summarize, we developed a (1+1)-dimensional
second-order viscous hydrodynamic model
with both shear and bulk viscosity to see the QGP dynamics
in the longitudinal direction.
There is no boost-invariance 
at both RHIC and LHC energies in the CGC initial conditions,
which causes the deformation of the entropy
per flow rapidity.
This indicates that the shapes of the (pseudo-)rapidity distributions
of hadrons observed in experiments
would reflect 
the initial gluon rapidity distributions
only indirectly
due to the hydrodynamic evolution.
This also motivates ones to correct the parameters,
in particular, $\lambda$ which controls the rapidity dependence
of entropy production,
in the initial conditions.
While the precise determination of the parameters 
should be left to a (3+1)-dimensional viscous 
hydrodynamic calculation, our current parameter settings 
suggest that a conventional
$\lambda$ which is determined without 
the hydrodynamic effect would be
smaller than the true $\lambda$ at RHIC but 
is not so much different at LHC around
the mid-rapidity region.
This could play an important role in explaining 
the gap between 
the CGC predictions of multiplicity based on RHIC data
and the latest experimental data at LHC \cite{Aamodt:2010pb}.
In future, 
more realistic wave functions under running-coupling
quantum evolution will be employed \cite{Albacete:2010ad}. 
We will discuss parameter dependences
and numerical aspects in detail in near future \cite{MH}.

\acknowledgments
\vspace*{-2mm}
The authors acknowledge fruitful discussion with Y.~Nara.
The work of A.M. was supported by JSPS Research Fellowships 
for Young Scientists.
The work of T.H. was partly supported by
Grant-in-Aid for Scientific Research
No.~22740151.  

%


\begin{thebibliography}{99}

\bibitem{Yagi:2005yb}
  K.~Yagi, T.~Hatsuda and Y.~Miake,
  Camb.\ Monogr.\ Part.\ Phys.\ Nucl.\ Phys.\ Cosmol.\  {\bf 23} (2005) 1.

\bibitem{Gelis:2010nm}
  F.~Gelis, E.~Iancu, J.~Jalilian-Marian and R.~Venugopalan,
  arXiv:1002.0333 [hep-ph].

\bibitem{Blaizot:2004px}
  J.~P.~Blaizot and F.~Gelis,
  Nucl.\ Phys.\  A {\bf 750} (2005) 148.


\bibitem{Kharzeev:2000ph}
D.~Kharzeev and M.~Nardi,
  Phys.\ Lett.\  B {\bf 507} (2001) 121.

\bibitem{Kharzeev:2001gp}
  D.~Kharzeev and E.~Levin,
  Phys.\ Lett.\  B {\bf 523} (2001) 79.


\bibitem{Kharzeev:2001yq}
  D.~Kharzeev, E.~Levin and M.~Nardi,
  Phys.\ Rev.\  C {\bf 71} (2005) 054903.


\bibitem{Kharzeev:2004if}
  D.~Kharzeev, E.~Levin and M.~Nardi,
  Nucl.\ Phys.\  A {\bf 747} (2005) 609.


\bibitem{Albacete:2007sm}
  J.~L.~Albacete,
  Phys.\ Rev.\ Lett.\  {\bf 99} (2007) 262301.

\bibitem{Levin:2010zy}
  E.~Levin and A.~H.~Rezaeian,
  Phys.\ Rev.\  D {\bf 82} (2010) 054003;

  E.~Levin and A.~H.~Rezaeian,
  arXiv:1011.3591 [hep-ph];

  E.~Levin and A.~H.~Rezaeian,
  arXiv:1102.2385 [hep-ph].

\bibitem{McLerran:2010ex}
  L.~McLerran and M.~Praszalowicz,
  Acta Phys.\ Polon.\  B {\bf 41} (2010) 1917.


\bibitem{HKH}
  P.~F.~Kolb, P.~Huovinen, U.~W.~Heinz and H.~Heiselberg,
  Phys.\ Lett.\  B {\bf 500} (2001) 232;

  P.~Huovinen, P.~F.~Kolb, U.~W.~Heinz, P.~V.~Ruuskanen and S.~A.~Voloshin,
  Phys.\ Lett.\  B {\bf 503} (2001) 58;

  P.~F.~Kolb, U.~W.~Heinz, P.~Huovinen, K.~J.~Eskola and K.~Tuominen,
  Nucl.\ Phys.\  A {\bf 696} (2001) 197.

\bibitem{Teaney:2000cw}
  D.~Teaney, J.~Lauret, and E.~V.~Shuryak,
  Phys.\ Rev.\ Lett.\  {\bf 86} (2001) 4783.

\bibitem{Hiranov2eta}
  T.~Hirano,
  Phys.\ Rev.\ C {\bf 65} (2002) 011901.

\bibitem{HT02}
  T.~Hirano and K.~Tsuda,
  Phys.\ Rev.\ C {\bf 66} (2002) 054905.


\bibitem{Song:2010mg}
  H.~Song, S.~A.~Bass, U.~W.~Heinz, T.~Hirano and C.~Shen,
  arXiv:1011.2783 [nucl-th].


\bibitem{Hirano:2004en}
  T.~Hirano and Y.~Nara,
  Nucl.\ Phys.\  A {\bf 743} (2004) 305.

\bibitem{hydroreviews}
  P.~Huovinen and P.~V.~Ruuskanen,
  Ann.\ Rev.\ Nucl.\ Part.\ Sci.\  {\bf 56}, (2006) 163;

  T.~Hirano, N.~van der Kolk and A.~Bilandzic,
  arXiv:0808.2684 [nucl-th];

  U.~W.~Heinz,
  arXiv:0901.4355 [nucl-th];

  P.~Romatschke,
  arXiv:0902.3663 [hep-ph];

  D.~A.~Teaney,
  arXiv:0905.2433 [nucl-th].
  
\bibitem{Bjorken:1982qr}
  J.~D.~Bjorken,
  Phys.\ Rev.\ D {\bf 27} (1983) 140.

\bibitem{GBW}
K.~Golec-Biernat and M.~Wusthoff,
Phys.\ Rev.\ D {\bf 59} (1999) 014017;
%

K.~Golec-Biernat and M.~Wusthoff,
Phys.\ Rev.\ D {\bf 60} (1999) 114023.


\bibitem{Chu:1986fu}
  M.~C.~Chu,
  Phys.\ Rev.\  D {\bf 34}  (1986) 2764.

\bibitem{Akase1989}
  Y.~Akase, S.~Dat\'{e}, M.~Mizutani, S.~Muroya, M.~Namiki and M.~Yasuda,
  Prog.\ Theor.\ Phys.\  {\bf 82} (1989) 591.
  
\bibitem{Bozek:2007qt}
  P.~Bozek,
  Phys.\ Rev.\  C {\bf 77} (2008) 034911.

\bibitem{Monnai:2010qp}
  A.~Monnai and T.~Hirano,
  Nucl.\ Phys.\  A {\bf 847} (2010) 283; 

  A.~Monnai and T.~Hirano,
  J. Phys.: Conf. Ser. {\bf270} (2011) 012042.
  
\bibitem{Israel:1979wp}
  W.~Israel and J.~M.~Stewart,
  Annals Phys.\  {\bf 118} (1979) 341.

\bibitem{PPM}
  P. Colella and P. R. Woodward, J. Comp. Phys. \textbf{54} (1984) 174.
  
 \bibitem{MH}
  A.~Monnai and T.~Hirano, in preparation.

\bibitem{Borsanyi:2010cj}
  S.~Borsanyi {\it et al.},
  JHEP {\bf 1011} (2010) 077.
  
\bibitem{Son_visc}

  P.~Kovtun, D. T.~Son, and A. O.~Starinets,
  Phys.\ Rev.\ Lett.\  {\bf 94} (2005) 111601.

\bibitem{Hosoya:1983id}
  A.~Hosoya, M.~a.~Sakagami, and M.~Takao,
  Annals Phys.\  {\bf 154} (1984) 229.

\bibitem{Drescher:2006ca}
  H.~J.~Drescher and Y.~Nara,
  Phys.\ Rev.\  C {\bf 75} (2007) 034905; 

  H.~J.~Drescher and Y.~Nara,
  Phys.\ Rev.\  C {\bf 76} (2007) 041903.

\bibitem{Kharzeev:2002ei}
D.~Kharzeev, E.~Levin and M.~Nardi,
  Nucl.\ Phys.\  A {\bf 730} (2004) 448
  [Erratum-ibid.\  A {\bf 743} (2004) 329].

\bibitem{Hirano:2009ah}
T.~Hirano and Y.~Nara,
Phys.\ Rev.\  C {\bf 79} (2009) 064904.
  
\bibitem{Morita:2002av}
  K.~Morita, S.~Muroya, C.~Nonaka and T.~Hirano,
  Phys.\ Rev.\  C {\bf 66} (2002) 054904.
  
\bibitem{Buchel:2007mf}
  A.~Buchel,
  Phys.\ Lett.\  B {\bf 663} (2008) 286.

\bibitem{Aamodt:2010pb}
  K.~Aamodt {\it et al.}  [The ALICE Collaboration],
  Phys.\ Rev.\ Lett.\  {\bf 105} (2010) 252301.

\bibitem{Albacete:2010ad}
  J.~L.~Albacete and A.~Dumitru,
  arXiv:1011.5161 [hep-ph].


\end{thebibliography}
\end{document}